\documentclass[a4paper,11pt,
colorlinks=true,urlcolor=blue,anchorcolor=blue,citecolor=blue,filecolor=blue,linkcolor=blue,menucolor=blue,linktocpage=true,pdfa=true]{article}
\usepackage{jinstpub} 

\usepackage{siunitx}

\usepackage{graphicx}
\usepackage{subcaption}
\usepackage{multirow}
\usepackage{booktabs, tabularx}

\definecolor{deepiris}{RGB}{93,63,211} 

\begin{document}

\vfill

\title{X-ray characterization of fully-depleted p-channel Skipper-CCDs for the DarkNESS mission}

\author[a,1]{Phoenix Alpine,\note{Corresponding author.}}
\author[b]{Ana M. Botti,}
\author[b]{Brenda A. Cervantes-Vergara,}
\author[b]{Claudio R. Chavez,}
\author[c]{Fernando Chierchie,}
\author[b,d,e]{Alex Drlica-Wagner,}
\author[f,b,d]{Juan Estrada,}
\author[g]{Erez Etzion,}
\author[a]{Michael Lembeck,}
\author[h]{Pilar L\'{o}pez Maggi,}
\author[d]{Joseph Noonan,}
\author[e]{Brandon Roach,}
\author[b,e,1]{Nathan Saffold,}
\author[b]{Javier Tiffenberg}

\affiliation[a]{University of Illinois, Urbana-Champaign, IL, 61801, USA}
\affiliation[b]{Fermi National Accelerator Laboratory, Batavia, IL 60510, USA}
\affiliation[c]{Instituto de Inv. en Ing. El\'ectrica ``Alfredo C. Desages'' (IIIE) CONICET, Bah\'ia Blanca, Argentina}
\affiliation[d]{Department of Astronomy and Astrophysics, University of Chicago, Chicago, IL 60637, USA}
\affiliation[e]{Kavli Institute for Cosmological Physics, University of Chicago, Chicago, IL 60637, USA}
\affiliation[f]{Instrumentation Division, Brookhaven National Laboratory, Upton, NY 11973, USA}
\affiliation[g]{School of Physics and Astronomy, Tel Aviv University, Tel Aviv, 69978, Israel}
\affiliation[h]{Universidad de Buenos Aires, Facultad de Ciencias Exactas y Naturales, Departamento de Física, Buenos Aires, Argentina}

\emailAdd{alpine2@illinois.edu}
\emailAdd{nsaffold@fnal.gov}

\abstract{
The Dark matter Nanosatellite Equipped with Skipper Sensors (DarkNESS) mission is a 6U CubeSat designed to search for X-ray lines from decaying dark matter using Skipper-CCDs.
Thick, fully-depleted p-channel Skipper-CCDs provide low readout noise and high quantum efficiency for 1--10~keV X-rays, but their X-ray performance has not yet been demonstrated in the space environment.
DarkNESS will operate in low-Earth orbit, where trapped protons induce displacement damage in the sensor that increases charge-transfer inefficiency and degrades the X-ray energy resolution. This work measures the X-ray line response of Skipper-CCDs before and after proton irradiation and quantifies the associated degradation. A sensor was exposed to 217~MeV protons at a fluence of $8.4\times10^{10}\,\mathrm{protons\,cm^{-2}}$, corresponding to a displacement-damage dose more than an order of magnitude above the three-year expectation for representative mid-inclination and Sun-synchronous low-Earth orbits. A $^{55}$Fe source was used to compare the energy resolution of the beam-exposed quadrant to adjacent unexposed quadrants and a non-irradiated reference sensor. 
These measurements provide a quantitative assessment of radiation-induced spectral degradation in Skipper-CCDs and enable an estimate of the end-of-life X-ray energy resolution expected for DarkNESS operation in low-Earth orbit.
}

\maketitle
\thispagestyle{empty}
\pagenumbering{arabic} 

\section{Introduction}
\label{intro}

The Dark matter Nanosatellite Equipped with Skipper Sensors (DarkNESS) mission will deploy a 6U CubeSat in low-Earth orbit (LEO) to search for dark matter using Skipper Charge-Coupled Devices (Skipper-CCDs). The mission targets two dark matter channels: electron recoils from strongly interacting sub-GeV dark matter and X-ray lines from decaying dark matter. For the X-ray channel, the instrument's energy resolution determines the width of the analysis window used to search for unidentified X-ray lines,
and improved resolution reduces the background integrated within each energy bin. 
The mission targets the Galactic Center to accumulate a $1$\,Ms exposure, using wide-field photometry to resolve statistical excesses in the X-ray band. In parallel with its science program, DarkNESS serves as a technology demonstrator, advancing the technology readiness of Skipper-CCDs for space-based applications~\cite{Rauscher_2022}.

DarkNESS employs four thick, fully-depleted p-channel Skipper-CCDs that provide sub-electron readout noise and high quantum efficiency in the 1--10~keV band. Previous studies of thick, fully-depleted p-channel Skipper-CCDs confirmed stable readout and charge transfer after 217\,MeV proton exposure to a fluence that exceeds the expected multi-year LEO dose~\cite{roach:2024ProtonIrradiation,vergara:2025JINST}. However, the impact of radiation-induced damage on the X-ray spectral performance of Skipper-CCDs has not been studied. This work reports the response of an irradiated Skipper-CCD exposed to $^{55}$Fe X-rays, and compares it to non-irradiated reference sensors. The analysis measures the detector gain and energy resolution, and quantifies the charge transfer inefficiency (CTI) due to radiation-induced traps. We use the measured spectral broadening and efficiency loss to model spectral degradation under representative LEO trapped-proton environments, establishing the end-of-life performance baseline for the DarkNESS decaying dark matter search. 

The remainder of this paper is organized as follows. Section~\ref{skipperCCD} summarizes the Skipper-CCD architecture and the DarkNESS detector configuration. Section~\ref{irradiation} describes the proton-irradiation campaign and the laboratory reference dose. Section~\ref{methods} outlines the X-ray measurement setup, data analysis, and event-reconstruction procedure. Section~\ref{results} presents the measured spectral response before and after irradiation and uses these results to project end-of-life performance in representative LEO trapped-proton environments. Conclusions and future work are provided in Section~\ref{futurework}.

\section{Skipper-CCD Technology}
\label{skipperCCD}

Skipper-CCDs are silicon imagers that achieve ultra-low readout noise through a floating-gate amplifier that enables repeated, non-destructive measurements of the charge in each pixel~\cite{moroni:2011skipperCCD}. By averaging $N$ samples of each pixel, the skipper amplifier can reduce the readout noise as $\sigma_N\propto1/\sqrt{N}$, enabling sub-electron noise performance~\cite{tiffenberg:2017skipper}. This sub-electron readout noise combined with low dark current has enabled world-leading sensitivity in rare-event searches (Ref.~\cite{adari:2025SENSEI} and references therein). These same characteristics make Skipper-CCDs well suited for a broad range of astronomical imaging applications~\cite{Marrufo_Villalpando_2024}, although their performance has not yet been demonstrated in the space radiation environment.

To investigate how radiation exposure affects the X-ray performance of Skipper-CCDs, we study prototype sensors fabricated for the Oscura experiment~\cite{vergara:2024Oscura}. These devices are fabricated on high-resistivity \textit{n}-type silicon and employ a buried \textit{p}-channel architecture developed at Lawrence Berkeley National Laboratory~\cite{holland:2003LBNL}, which provides improved radiation tolerance compared to \textit{n}-channel designs~\cite{bebek:2002RadiationPchannel}. For X-ray spectroscopy, sensor thickness and depletion depth are key design parameters that determine the detector's energy response. The Oscura prototype devices can be fully depleted over a thickness of 725~$\mu$m, providing an X-ray absorption probability $\gtrsim$99\% at 10~keV.
The specifications and operating parameters for the Oscura prototype devices are summarized in Table~\ref{tab:sensor_summary}, with DarkNESS flight configurations indicated for comparison. 

\begin{table}[h]
\centering
\caption{Sensor Specifications and Operating Parameters}
\label{tab:sensor_summary}
\footnotesize
\setlength{\tabcolsep}{4pt}
\begin{tabularx}{\linewidth}{
    >{\raggedright\arraybackslash}X l
    >{\raggedright\arraybackslash}X l
}
\toprule
\multicolumn{2}{c}{\textbf{Sensor parameters}} & \multicolumn{2}{c}{\textbf{Operating parameters}} \\
\cmidrule(r){1-2}\cmidrule(l){3-4}
\textbf{Parameter} & \textbf{Value} & \textbf{Parameter} & \textbf{Value} \\
\midrule
Sensor type & Buried \textit{p}-channel & Operating temperature & 153/(170)\textsuperscript{\textasteriskcentered} K \\
Substrate & High-res \textit{n}-type & Bias voltage & 70/(40)\textsuperscript{\textasteriskcentered} V \\
Format & $1278\times1058$ & Depletion & Fully depleted \\
Pixel pitch & $15~\mu$m & Single-sample readout noise & $<4~e^-$ RMS \\
Readout & 4 Skipper amps & Entrance window\textsuperscript{\textasteriskcentered} & 50 nm Al \\
Si thickness & 725/(500)\textsuperscript{\textasteriskcentered}~$\mu$m & & \\
\bottomrule
\multicolumn{4}{l}{\scriptsize \textsuperscript{\textasteriskcentered} \textit{DarkNESS configuration}} \\
\end{tabularx}
\end{table}

While the Oscura prototype sensors form the basis of this study, the devices planned for the DarkNESS mission differ in several respects. The DarkNESS flight sensors incorporate a 50~nm aluminum entrance window to suppress optical and near-infrared loading in LEO while preserving high transmission of X-rays~\cite{botti_2026}. In addition, the silicon bulk is thinned from $725~\mu$m to 500~$\mu$m to reduce the number of pixels affected by charged-particle tracks during LEO operation~\cite{alpine:2025ASR}. Although there are several mission-driven design differences, the Oscura prototype sensors provide a suitable proxy for the DarkNESS flight devices to study radiation-induced changes in X-ray spectral performance.

\section{Proton Irradiation}
\label{irradiation}

Skipper-CCDs were irradiated at the Northwestern Medicine Proton Center in August 2023 using a clinical 217\,MeV proton therapy beam to study the impact of proton-induced displacement damage on detector performance. The laboratory beam provides a monoenergetic, high-energy proton exposure that serves as a controlled reference for interpreting radiation-induced degradation expected during multi-year operation in LEO.

\begin{figure}[h]
\centering
\includegraphics[width=0.85\textwidth]{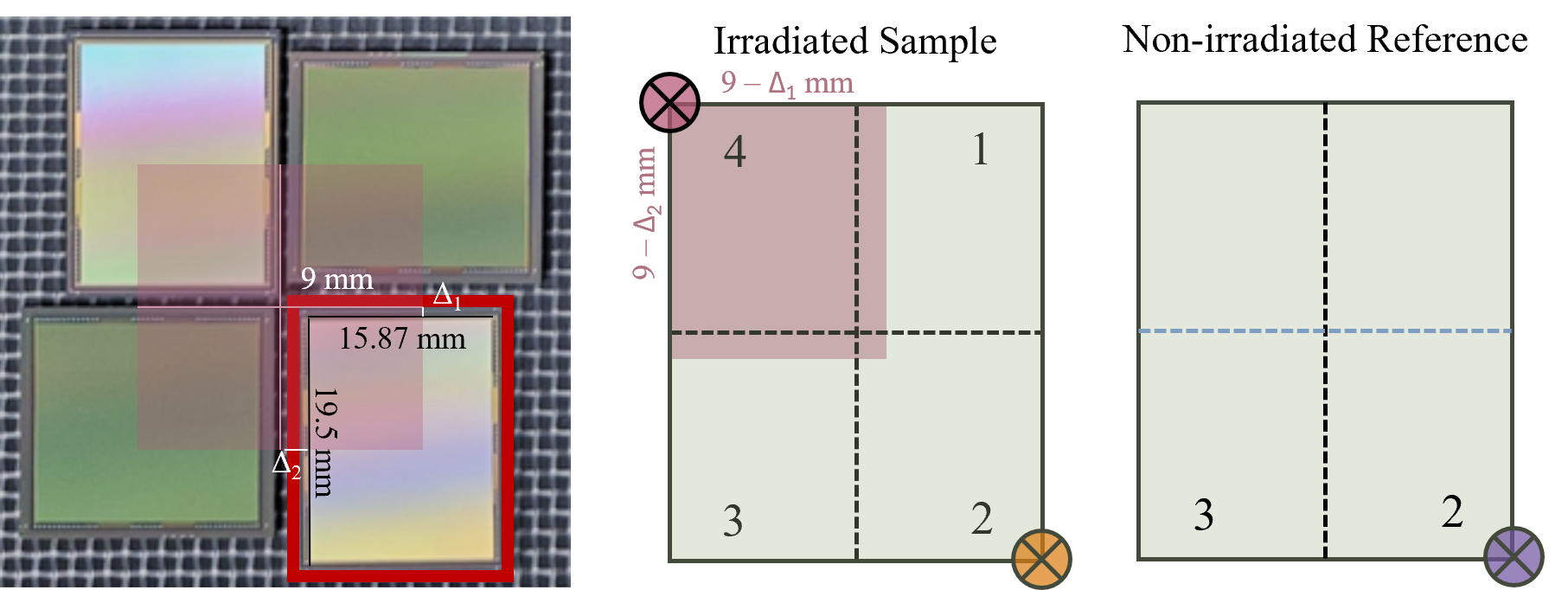}
\caption{
Sample arrangement during irradiation showing four Skipper-CCDs with the beam profile center aligned at the sensor carrier. The CCD under study (red outline) contains four skipper readout amplifiers, one in each corner of the chip. The square-profile 217~MeV proton beam (rose overlay) delivers a fluence of \mbox{$8.4\times10^{10}\,\mathrm{p\,cm^{-2}}$}, corresponding to decades of equivalent LEO exposure (18--42 years depending on orbit assumptions). Data from the irradiated adjacent amplifier~2 (IA2) of the irradiated CCD marked orange are compared with the equivalent amplifier~2 region of a non-irradiated (NI2) reference CCD marked iris. The beam-irradiated quadrant read out through amplifier~4 (BI4) provides the direct comparison region, marked in red.
}
\label{fig:configuration}
\end{figure}

The sensor analyzed in this work was irradiated alongside a set of Skipper-CCDs under the conditions described in Refs.~\cite{roach:2024ProtonIrradiation,vergara:2025ProtonDamage}, with the sensors held at room temperature and unbiased during exposure. The devices were mounted in a $2\times2$ array such that the beam intersected the central point between the four sensors (Fig.~\ref{fig:configuration}). Each sensor is read out by four independent amplifiers located at the corners of the device, providing separate readout of the four quadrants. In this configuration, radiation damage can affect both the charge-transfer properties of a quadrant and the response of its associated readout amplifier, requiring both effects to be evaluated independently. We label each quadrant with a two-letter prefix indicating the irradiation condition (BI: beam-irradiated, IA: irradiated-adjacent, NI: non-irradiated), followed by the amplifier number (1–4).

For the irradiated sensor studied here, quadrant~4 (BI4) received the full beam fluence, while an adjacent quadrant~2 (IA2) provides the primary on-chip reference region. Although the beam footprint partially overlapped neighboring quadrants, only the amplifier associated with quadrant~4 lay within the region of highest fluence, while the remaining amplifiers remained outside the beam core. A separate non-irradiated prototype Skipper-CCD was also studied, with quadrant~2 (NI2) selected for comparison. The irradiated quadrant received a fluence of $8.4\times10^{10}~\mathrm{p\,cm^{-2}}$, corresponding to a displacement-damage dose of $1.6\times10^{8}~\mathrm{MeV\,g^{-1}}$ (Si).

The measured displacement-damage dose (DDD) serves as the laboratory anchor for interpreting radiation-induced performance degradation in orbit. For a monoenergetic proton beam, the fluence--dose relation directly calibrates the effective Non-Ionizing Energy Loss (NIEL) at 217~MeV through $\mathrm{DDD}=\Phi\,\mathrm{NIEL}(E)$~\cite{Summers_1994}. This calibration is used to fold trapped-proton spectra from the AP9 model~\cite{Ginet_AP9} into predicted displacement-damage dose rates for representative DarkNESS orbits under conservative solar-minimum conditions. The resulting orbit-equivalent exposures correspond to approximately 40~years in an International Space Station (ISS) orbit at 410~km and $51.6^\circ$ inclination, and approximately 20~years in a sun-synchronous orbit (SSO) at 500~km and $97.0^\circ$ inclination. These representative cases bracket the range of radiation environments relevant for the DarkNESS mission.

\section{Methods}
\label{methods}
\subsection{Experimental Setup}
Energy resolution measurements were performed in a vacuum chamber operated below $10^{-4}$\,Torr. The Skipper-CCD was mounted in a copper fixture that was thermally anchored to a custom-machined temperature control plate. This plate was directly coupled to the cold tip of an Edwards PCC Compact Cryocooler, which provided the primary cooling interface. Temperature stabilization at 153\,K was maintained by a resistive heater and a resistive temperature detector embedded in the plate. The heater and temperature sensor were connected through vacuum feedthroughs to a LakeShore temperature controller, which provided closed-loop feedback for thermal stability throughout the data acquisition run. A ${}^{55}$Fe calibration source was affixed to the interior wall of the vacuum chamber, positioned a few centimeters from the detector's front surface. Detector biasing and clocking voltages were supplied by the Low Threshold Acquisition (LTA) electronics system~\cite{cancelo_2020}, with a bias voltage of $70$\,V applied to the back surface of the Skipper-CCD to deplete the silicon bulk. The LTA system interfaced with the control computer via Ethernet, and image data were recorded in FITS format for downstream processing.

\subsection{Image Processing}
\label{sec:imageprocessing}
All datasets were processed with a Python-based pipeline that performed skipper-sample averaging and baseline subtraction on each quadrant. Each raw image contains ten non-destructive samples per pixel; these samples were reshaped into per-pixel stacks and averaged to suppress electronic readout noise prior to event extraction. The overscan pixel distributions provide a measure of the readout noise; however, some overscan pixels contain charge due to energy depositions in the serial register during readout that can bias the noise estimation. To obtain a robust estimate of the read noise, we computed the median and median absolute deviation (MAD) of the overscan pixel values, masked pixels deviating by more than three MADs from the median, and converted the MAD of the remaining pixels to a Gaussian-equivalent standard deviation. The unmasked overscan pixels defined both the per-pixel noise ($\sigma_{\rm read}$) for each image and the pedestal level used for baseline correction. The image pedestal is subtracted row by row, and then the median of each column is subtracted column by column to correct residual column-dependent offsets introduced by vertical-shift transients. This preprocessing produces a bias- and baseline-corrected image that is used for X-ray event extraction.

\subsection{X-ray Extraction}
\label{sec:extraction}
X-ray events were extracted using a thresholded, connected-component algorithm applied to the processed images. A high event threshold (9$\sigma_{\rm read}$) seeded clusters, and a lower split threshold (3$\sigma_{\rm read}$) collected neighboring pixels sharing charge from the same interaction. Here, $\sigma_{\mathrm{read}}$ is estimated from the overscan distribution after skipper-sample averaging (Sec.~\ref{sec:imageprocessing}) and represents the effective per-pixel noise used for event detection. Typical noise levels are approximately 3.5~$e^{-}$ in the non-irradiated quadrants and 4.7~$e^{-}$ in the irradiated quadrant, reflecting contributions from averaged readout noise and spurious charge generated during readout.
Pixels above the split threshold were grouped using 4-neighbor connectivity to reconstruct the compact geometries expected from keV-scale X-ray absorption. For each connected region, cluster properties (including total energy, peak pixel energy, energy-weighted centroid, pixel multiplicity, and bounding-box dimensions) were calculated directly from the pixel values in analog-to-digital units (ADU).

\begin{figure}[t]
\centering
\includegraphics[width=1\textwidth]{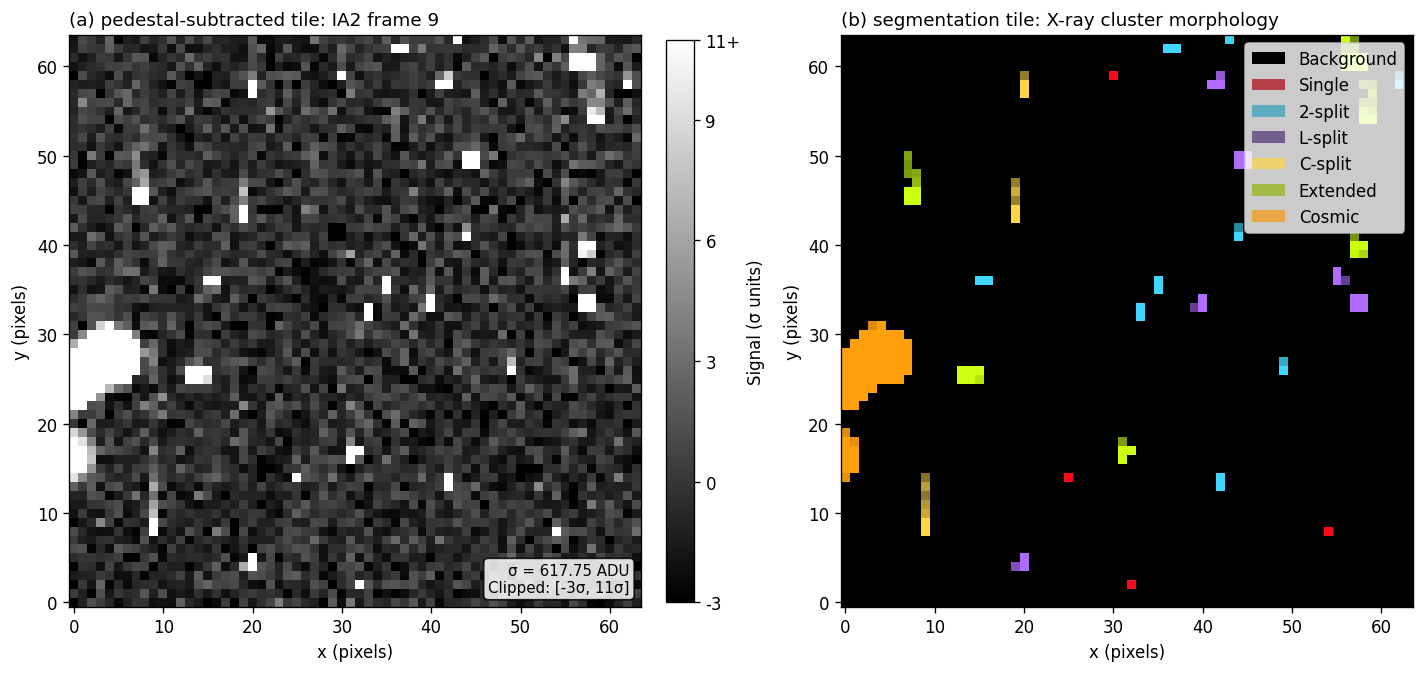}
\caption{Thresholded 4-neighbor connected-component clustering. (Left) A $64 \times 64$ pixel tile of the pedestal-subtracted image with values in units of readout noise ($\sigma_{\rm read}$). The scale highlights pixels above the split threshold ($3\sigma_{\rm read}$) used to collect charge and the event threshold ($9\sigma_{\rm read}$) used to seed clusters. (Right) The corresponding segmentation mask. Events are classified by topological grade, including charged-particle tracks (orange), single-pixel (red), 2-split (cyan), L-split (purple), and extended (green). The C-split grade (yellow) identifies column-wise trails from vertical CTI.}
\label{fig:morphology}
\end{figure}

Cluster morphology was classified with a topology-based grade scheme, shown in Fig.~\ref{fig:morphology}. “Single,” “2-split,” and “L-split” grades denote compact charge packets occupying one or a few adjacent pixels. A “C-split” grade identifies short, column-aligned sequences produced by deferred charge from charge-transfer inefficiency; these appear as three or more vertically aligned pixels with minimal lateral extent. Regions with large pixel counts, pronounced elongation, or irregular shapes were labeled “extended.” Extended clusters (cosmic-ray tracks and long ionization trails) were removed using cuts on pixel count and aspect ratio. The validated event catalog contains the clustered Mn--K$\alpha$ and K$\beta$ interactions used for spectroscopic analysis.

\subsection{Spectral measurements}
To assess the gain and energy resolution of the Skipper-CCDs, we constructed a histogram of the cluster energies in ADU, selecting clusters that passed the morphology criteria described above. The calibration is performed using the Mn--K$\alpha$ (5.895\,keV) and Mn--K$\beta$ (6.49\,keV) lines. The two lines were fit with a double-Gaussian model with the centroid ratio and relative amplitude fixed to the known Mn K$\alpha$/K$\beta$ transition energies and branching ratios. The Gaussian width provided the instrumental energy resolution at 5.895~keV, and the centroid separation ($\Delta E_{\beta-\alpha}$) served as an energy-scale check. The measured K$\alpha$ variance was decomposed into contributions from Fano statistics, electronic read noise, and a residual broadening term attributed to radiation-induced charge-transfer losses. Spatial uniformity was assessed by tracking the raw-pixel flux (in ADU) as a function of column index to identify gain variations along the vertical transfer direction. The resulting gain, full width at half maximum (FWHM) energy resolution, and residual broadening for each amplifier region form the laboratory inputs to the DarkNESS performance models.

\section{Results}
\label{results}
\subsection{Gain, Energy Resolution, and Charge Transfer Performance}
\label{sec:spectra}
\begin{figure}[t]
\centering
\includegraphics[width=1\textwidth]{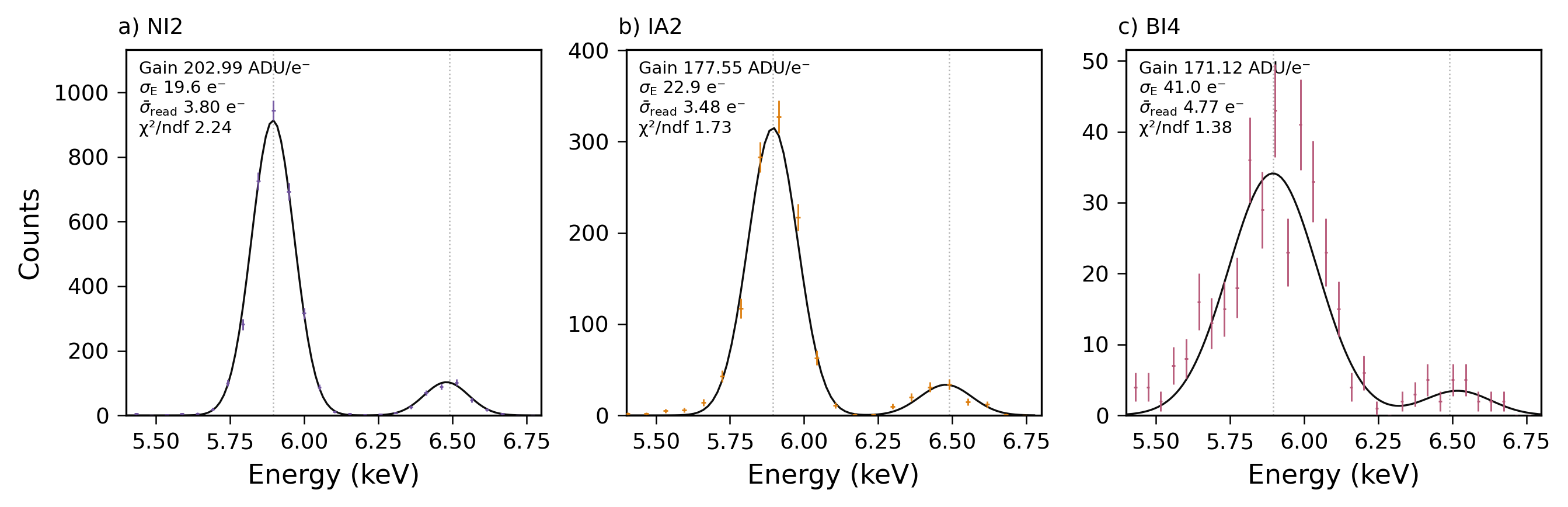}
\caption{
Full-quadrant $^{55}$Fe X-ray spectra for the non-irradiated reference quadrant (NI2), the adjacent quadrant of the irradiated sensor (IA2), and the beam-exposed quadrant (BI4). Points with cross-mark error bars show the measured cluster-energy distributions, and solid curves give the best-fit double-Gaussian model to the Mn--K$\alpha$ and Mn--K$\beta$ lines. The vertical dashed lines mark the nominal Mn--K$\alpha$ (5.895~keV) and Mn--K$\beta$ (6.490~keV) peak energies. Relative to NI2, IA2 exhibits only modest broadening of the Mn--K$\alpha$ line, whereas BI4 shows a factor-of-two increase in FWHM and pronounced low-energy tailing associated with radiation-induced displacement damage.}
\label{fig:histograms}
\end{figure}

To quantify the impact of radiation damage on spectral performance, we compare the $^{55}$Fe energy resolution of the quadrant that was irradiated in the proton beam (BI4) with the non-irradiated and irradiated-adjacent reference quadrants (NI2 and IA2). Figure~\ref{fig:histograms} shows the full-quadrant spectra, and Table~\ref{tab:result} summarizes the corresponding gain and energy resolution parameters. Each quadrant's performance is also evaluated using a separate dataset of 100-row images that restricts the active area to pixels near the serial register, thereby minimizing contributions from vertical charge-transfer inefficiency (CTI). 
While the amplifier gain varies by $\sim$10--15\% between quadrants (a level consistent with typical quadrant-to-quadrant gain differences), the 100-row datasets recover a Mn-K$\beta$--K$\alpha$ centroid separation that agrees with the theoretical 595~eV spacing to within a small fraction of the instrumental resolution ($<$0.1 FWHM).
Furthermore, the irradiated quadrant (BI4) improves from a distorted separation of 622.9~eV in the full quadrant to 600.7~eV in the 100-row dataset, suggesting that the spectral degradation observed in the full-quadrant data arises from charge-transfer effects.

\begin{table}[htbp]
\centering
\caption{Spectral Calibration and Resolution}
\label{tab:result}
\begin{tabular*}{\textwidth}{l@{\extracolsep{\fill}}cccc}
\toprule
\textbf{Sample} & \textbf{Gain [ADU/e$^-$]} & \textbf{FWHM [eV]} & $\Delta E_{\beta-\alpha}$ \textbf{[eV]} & $\chi^2$/ndf \\
\midrule
\multicolumn{5}{c}{\textit{full quadrant exposures}} \\
NI2          & $202.99 \pm 0.02$ & $169 \pm 1$ & 585.1 & 248.1/111 (2.24) \\
IA2          & $177.55 \pm 0.04$ & $197 \pm 2$ & 583.7 & 152.6/88 (1.73) \\
BI4          & $171.12 \pm 0.14$ & $352 \pm 9$ & 622.9 & 196.9/143 (1.38) \\
\midrule
\multicolumn{5}{c}{\textit{100-row exposures}} \\
NI2          & $203.42 \pm 0.04$ & $167 \pm 2$ & 587.7 & 99.8/92 (1.08) \\
IA2          & $178.48 \pm 0.05$ & $181 \pm 3$ & 579.0 & 100.4/92 (1.09) \\
BI4          & $172.80 \pm 0.14$ & $250 \pm 9$ & 600.7 & 95.8/99 (0.97) \\
\bottomrule
\end{tabular*}
\end{table}

\begin{figure}[t]
  \centering
  \includegraphics[width=1\textwidth]{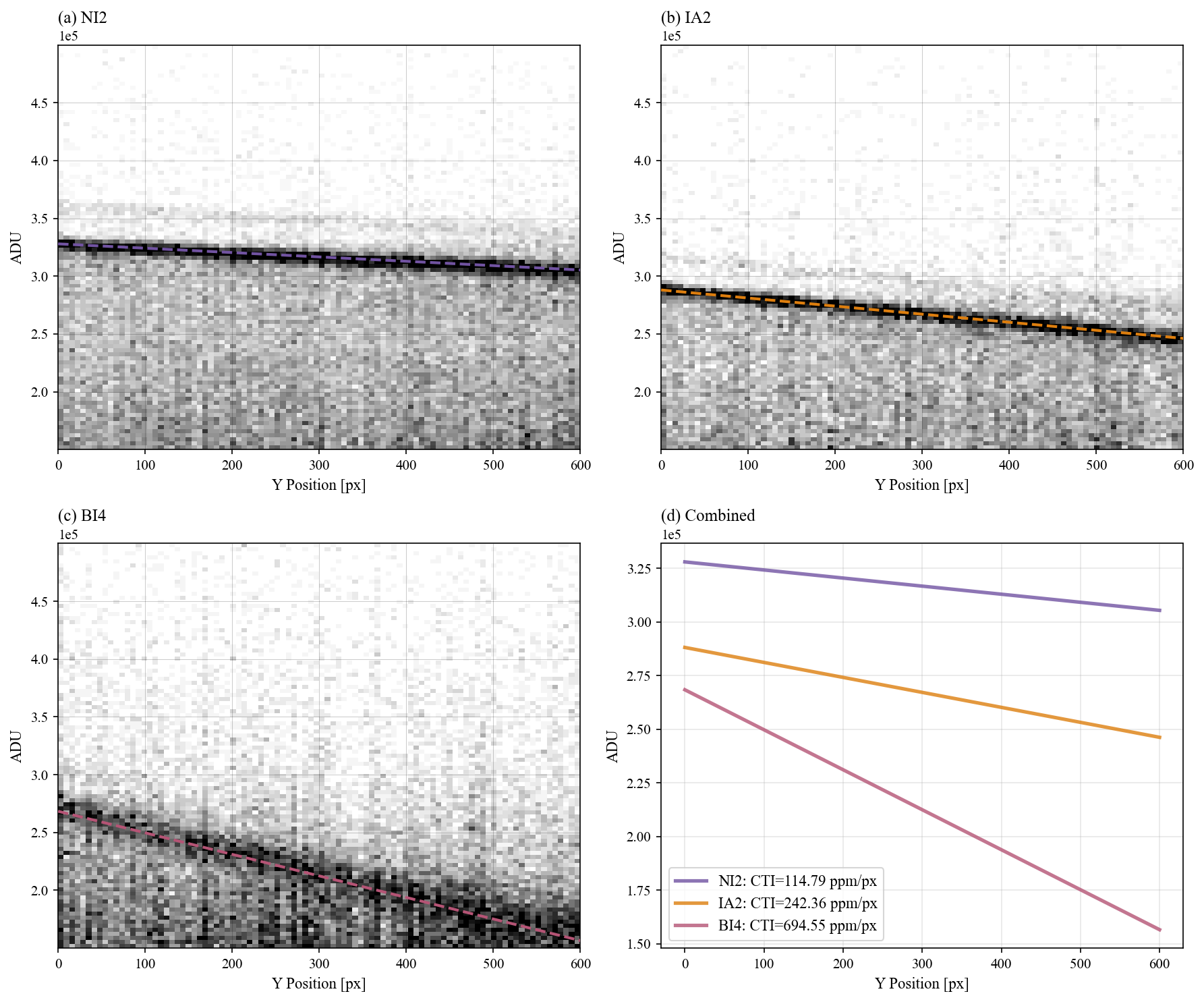}
  \caption{
  Pixel-value distributions as a function of vertical transfer distance (Y-position) for NI2 (top-left), IA2 (top-right), and BI4 (bottom-left), highlighting the Mn–K$\alpha$ (5.9 keV) signal band.
  The bottom-right panel overlays the ridge fits used to track the Mn–K$\alpha$ signal through the distributions. NI2 and IA2 show shallow, nearly parallel slopes, indicating minimal CTI, while BI4 exhibits a steeper decline consistent with increased CTI in the irradiated quadrant.
  }
  \label{fig:transfer}
\end{figure}

Examining the two non-irradiated quadrants, IA2 exhibits a modest but measurable broadening relative to NI2, with $\sim$13\% larger FWHM for the Mn--K$\alpha$ line in the full-quadrant dataset. IA2 also shows a slightly steeper vertical-transfer slope than NI2 (Fig.~\ref{fig:transfer}), consistent with a small increase in parallel CTI. Since parallel CTI increases with the number of charge transfers, it primarily affects full-quadrant images and has less influence on 100-row images, which sample only pixels near the serial register. Accordingly, the IA2 100-row FWHM is substantially closer to NI2.

The irradiated quadrant BI4 shows a qualitatively different behavior. In the full-quadrant data, BI4 exhibits a substantially broadened Mn–K$\alpha$ peak ($352\pm9$ eV), far exceeding the 170--200\,eV widths in NI2 and IA2. The modest increase in effective per-pixel noise in the irradiated quadrant (Sec.~\ref{sec:extraction}) is insufficient to account for the observed spectral broadening. The BI4 spectrum also exhibits a pronounced low-energy tail (Fig.~\ref{fig:histograms}), which indicates energy loss during charge transfer. This behavior correlates with the substantially higher trap density measured in the irradiated region ($\sim$$4\times 10^{-2}$ pix$^{-1}$), more than an order of magnitude larger than in IA2~\cite{vergara:2025ProtonDamage}. Restricting the analysis to the 100-row BI4 dataset reduces the fitted FWHM from $352\pm9$~eV to $250\pm9$~eV, which is consistent with trap-induced charge-transfer losses. The strong dependence on the transfer distance indicates that vertical CTI, rather than changes in amplifier gain or electronic noise, is the dominant source of spectral degradation in the irradiated quadrant.

To quantify the CTI in each quadrant, we construct a vertical charge-transfer curve by tracking the characteristic pixel signal associated with the Mn--K$\alpha$ (5.9~keV) peak as a function of vertical transfer distance (Y-position), extracted from ridge fits to the pixel-value distributions.
The resulting curves, shown in Fig.~\ref{fig:transfer}, further support the interpretation that the observed spectral degradation is driven by increased CTI.
NI2 and IA2 display shallow and nearly parallel trends as a function of Y-position, indicating minimal charge loss across the transfer length. In contrast, BI4 exhibits a significantly steeper slope with transfer distance, consistent with increased CTI. Linear fits to these trends yield a BI4 CTI approximately six times that of NI2, while IA2 remains within a factor of two of the non-irradiated reference. The larger CTI observed in BI4 is consistent with the elevated trap densities measured in Ref.~\cite{vergara:2025ProtonDamage}, indicating that the dominant radiation damage is confined to the beam-exposed region and manifests primarily as increased CTI.

Increased CTI in the irradiated quadrant also leads to a redistribution of event morphologies through grade morphing~\cite{Townsley_2002}. During charge transfer, radiation-induced traps capture and subsequently release charge, producing deferred charge that appears as trailing pixels. For the clocking times employed here, this trailing structure is predominantly aligned along the column direction, although different clocking schemes would alter the trailing morphology. As a result, compact X-ray events that would be classified as single- or few-pixel clusters in non-irradiated regions can be reclassified as C-split or extended clusters in the presence of elevated CTI. This effect is reflected in Table~\ref{tab:morphology}, which shows a substantial suppression of compact event grades and a corresponding increase in non-compact morphologies in BI4 relative to both NI2 and the on-chip reference quadrant IA2.

Because the event-selection criteria reject extended clusters, grade morphing reduces the effective detection efficiency for X-ray events. To quantify this effect, we define a grade retention efficiency (GRE) as the ratio of accepted compact X-ray events (single, 2-split, and L-split) in a given region relative to a reference. Using the 100-row dataset, BI4 retains only $\sim$26\% of the compact event population observed in NI2, and $\sim$35\% relative to IA2, indicating that radiation-induced charge trailing leads to an efficiency loss of reconstructable X-ray events beyond that expected from modest on-chip CTI alone.

\begin{table}[h]
\centering
\caption{Cluster Morphology Distribution (per image)}
\label{tab:morphology}
\begin{tabular*}{\linewidth}{@{\extracolsep{\fill}}lccccc}
\toprule
\textbf{Sample} & \textbf{Single} & \textbf{2-Split} & \textbf{L-Split} & \textbf{C-Split} & \textbf{Extended} \\
\midrule
\multicolumn{6}{c}{\textit{100-row exposures}} \\
NI2          & 29 & 58  & 68  & 16  & 140 \\
IA2          & 20 & 40  & 49  & 20  & 156 \\
BI4          & 1  & 11  & 29  & 3   & 193 \\
\bottomrule
\end{tabular*}
\end{table}

\begin{figure}[t]
  \centering
  \includegraphics[width=1\textwidth]{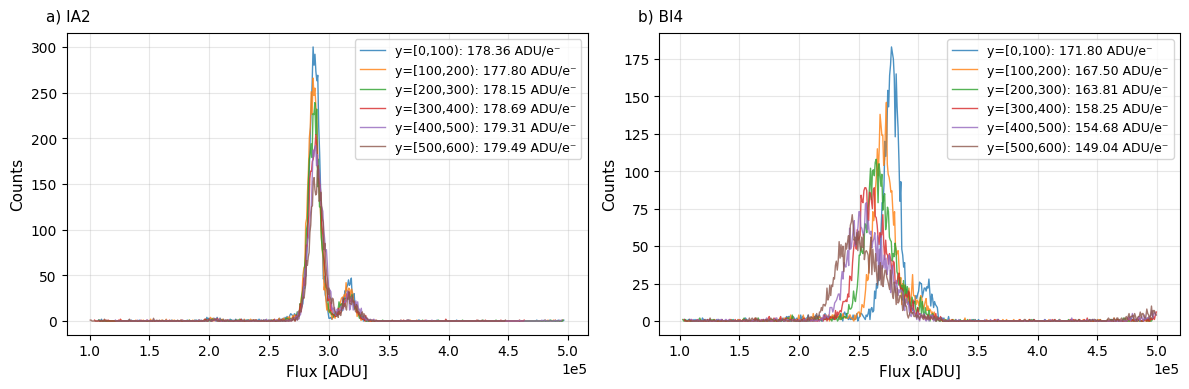}
  \caption{Mn--K$\alpha$ spectra for vertically segmented, 100-row subsets in NI2, IA2, and BI4. 
  The BI4 segments display broadened peaks with enhanced low-energy tailing relative to the nearly 
  identical, symmetric line shapes in NI2 and IA2, confirming that the observed broadening is intrinsic 
  to the irradiated segment rather than an artifact of quadrant size or sampling.}
  \label{fig:segmented_spectra}
\end{figure}

Figure~\ref{fig:segmented_spectra} presents the Mn--K$\alpha$ spectra for 100-row slices of the active area in IA2 and BI4. The segmented spectra illustrate that BI4 retains a broader K$\alpha$ peak with increasing low-energy tailing even when restricted to a common transfer length, whereas IA2 remains symmetric. This comparison reveals that the observed broadening is not an artifact of gain variation or amplifier degradation, but reflects intrinsic charge-transfer and collection losses in the irradiated segment.

Collectively, the gain, energy resolution, and CTI measurements provide a consistent, quantitative picture of displacement-damage effects in the irradiated quadrant. The increase from $\sim$169~eV to $352$~eV FWHM in BI4 for the full-quadrant exposure corresponds to a $\sim$183~eV degradation for a laboratory displacement-damage dose of $1.6\times10^{8}~\mathrm{MeV\,g^{-1}}$. In the following section, we use this empirical broadening as a first-order calibration relating displacement-damage dose to energy-resolution degradation, allowing the LEO trapped-proton environment modeled with AP9 to be mapped to an estimated end-of-mission FWHM.

\subsection{Projected Mission Performance}
\label{sec:radiation}

Laboratory irradiation provides an empirical anchor for predicting detector performance in the LEO radiation environment. We compute the expected degradation by scaling the measured 217~MeV damage to the broad trapped-proton spectrum modeled with AP9. The analysis links the laboratory DDD to on-orbit exposure using the NIEL function for silicon to scale the curve as $E^{-0.25}$ for proton energies $E < 300$~MeV, and normalizing to match the measured damage coefficient at the 217~MeV anchor point. This accounts for the greater NIEL-weighted damage per proton at lower energies, which dominate the South Atlantic Anomaly.

The trapped-proton environment is simulated with AE9/AP9-IRENE (v1.58) using a Perturbed Mean scheme. Flux rates are established using representative orbital sampling, integrating proton energies from 10 MeV to 316 MeV to encompass the primary displacement damage range. A conservative design limit is obtained from 40 Monte Carlo-sampled Perturbed Mean scenarios by calculating the cumulative displacement damage dose for each scenario and selecting the 95th percentile of the resulting distribution. Two reference orbits are considered: an ISS-like mid-inclination orbit ($51.6^\circ$, 410~km) and a higher-radiation Sun-synchronous orbit (SSO; $97^\circ$, 550~km, 12{:}00 LTAN) due to reduced geomagnetic shielding.

End-of-life energy resolution is projected using two accumulation models derived from the beam data. The linear model treats displacement damage as an additive contribution to the peak width, consistent with the linear scaling of CTI with proton fluence observed in p-channel CCDs~\cite{bebek:2002RadiationPchannel}. It is parameterized by the empirical damage coefficient:
\begin{equation}
    k_{\mathrm{DDD}} \equiv
    \frac{\Delta \mathrm{FWHM}}{\Delta \mathrm{DDD}}
    \approx 1.1\times10^{-6}\,
    \mathrm{eV}\,(\mathrm{MeV\,g^{-1}})^{-1},
    \label{eq:kddd}
\end{equation}
where $\Delta$FWHM is referenced to the non-irradiated energy resolution measured in NI2. This coefficient provides a conservative calibration for projecting in-orbit degradation. A quadrature model assumes radiation-induced broadening adds as an independent variance term, representing the stochastic fluctuations associated with charge transfer~\cite{janesick2001scientific}:
\begin{equation}
    \mathrm{FWHM}_{\mathrm{quad}}(t) = \sqrt{\mathrm{FWHM}_0^2 + (k_{\mathrm{DDD}} \cdot \mathrm{DDD}(t))^2},
    \label{eq:fwhm_quad}
\end{equation}
where $\mathrm{FWHM}_0$ is the non-irradiated energy resolution measured in the reference region (NI2).
The quadrature model reflects the physical summation of uncorrelated variance terms (e.g., Fano statistics, read noise, and trap-induced dispersion). These models are anchored to a single irradiation dose to provide an empirical normalization for mapping laboratory displacement damage to on-orbit exposure. Additional irradiations at intermediate fluences will be required to establish the detailed functional dependence of spectral degradation on dose.

For the DarkNESS X-ray line search, the instrument detects the signal as a statistical flux excess above the modeled astrophysical background, distinguishing the dark matter halo from thermal emission via spatial modulation (on-halo vs.\ off-halo pointing). In this context, the energy resolution primarily determines the continuum background integrated within the signal energy window, motivating constraints on radiation-induced spectral degradation.
Under the worst-case SSO scenario (3-year duration), the conservative linear model predicts a broadening to 198~eV (Table~\ref{tab:leo_degradation}). This corresponds to a $\sim$17\% increase in bin width relative to the non-irradiated detector. While this slightly increases the integrated background counts per bin, it does not significantly degrade the signal-to-noise ratio. The instrument's high grasp ($\sim$4.6~cm$^2$\,sr) and exposure ($\sim$1~Ms) provide sufficient statistical precision to resolve the residual flux excess despite this modest increase in energy resolution. The analysis confirms that the Skipper-CCD architecture maintains the necessary spectral stability for the DarkNESS photometric detection strategy in the LEO radiation environment.

\begin{table}[t]
\centering
\caption{Predicted spectral degradation (AP9 Perturbed Mean, 95\% Confidence)}
\label{tab:leo_degradation}
\begin{tabular*}{\textwidth}{l@{\extracolsep{\fill}}ccccc}
\toprule
Orbit & Duration & $\Phi_{95}$ [p~cm$^{-2}$] & DDD$_{95}$ [MeV~g$^{-1}$] & FWHM$_{\mathrm{lin}}$ [eV] & FWHM$_{\mathrm{quad}}$ [eV] \\
\midrule
ISS ($51.6^\circ$) & 1 yr & $1.29\times 10^{9}$ & $3.78\times 10^{6}$ & 173.32 & 169.06 \\
ISS ($51.6^\circ$) & 3 yr & $3.88\times 10^{9}$ & $1.13\times 10^{7}$ & 181.96 & 169.50 \\
SSO ($97.0^\circ$) & 1 yr & $2.89\times 10^{9}$ & $8.58\times 10^{6}$ & 178.82 & 169.28 \\
SSO ($97.0^\circ$) & 3 yr & $8.66\times 10^{9}$ & $2.57\times 10^{7}$ & 198.45 & 171.55 \\
\bottomrule
\end{tabular*}
\end{table}

\section{Conclusions and Future Work}
\label{futurework}
This work presents a laboratory study of radiation-induced degradation in the X-ray spectral performance of thick, fully depleted p-channel Skipper-CCDs. Using proton irradiation at 217~MeV and $^{55}$Fe X-ray spectroscopy, we compared the response of a beam-exposed quadrant to adjacent unirradiated regions and to a non-irradiated reference sensor. While the skipper amplifier remained operational after irradiation, these measurements demonstrate that radiation damage produces trap-induced charge-transfer inefficiency, which results in spectral broadening, low-energy tailing, and a reduction in reconstructable X-ray events. By anchoring the measured degradation to a calibrated displacement-damage dose and mapping it to trapped-proton environments predicted by AP9, we established an empirical estimate of the end-of-life X-ray energy resolution expected for DarkNESS operation in low-Earth orbit. The irradiation campaign presented here was performed at room temperature with the sensors unbiased; future irradiation studies will extend this work to cold, biased operating conditions representative of in-orbit DarkNESS operation.

Future work extends this X-ray characterization to DarkNESS flight-configured hardware currently in fabrication. The flight unit consists of a four-sensor Skipper-CCD module incorporating the aluminum optical entrance window, flight thermal interface, and the NanoSatellite implementation of the Low-Threshold Acquisition (sLTA) electronics, as described in Ref.~\cite{alpine:2025ASR}. Characterization of this integrated module establishes a mission-relevant benchmark of detector response, gain, and charge-transfer performance under the sLTA-specific clocking, biasing, and thermal conditions planned for on-orbit operation.

In parallel, continued study focuses on the development of improved event-clustering and reconstruction algorithms. The current connected-component approach, which relies on fixed morphological thresholds, exhibits limited performance in disentangling pile-up and overlapping events, particularly in the presence of CTI-induced deferred-charge morphologies that generate extended charge trails. These limitations are most evident in the BI4 dataset, where increased vertical CTI produces long, asymmetric charge tails that distort event geometries. Consequently, some compact X-ray interactions can be misclassified as extended or C-split events, reducing the reconstruction efficiency and degrading spectral fidelity.

Additional irradiation campaigns at fluences corresponding to the expected three-year end-of-mission exposure provide a means to assess clustering robustness under representative LEO damage conditions. Dedicated reconstruction strategies targeting deferred-charge morphologies are also explored to recover misclassified X-ray events while maintaining effective background rejection. 

\section*{Acknowledgments}
We are grateful to the staff of the Northwestern Medicine Proton Center for support during beam tests. The fully depleted Skipper-CCD was developed at Lawrence Berkeley National Laboratory, as were the designs described in this work. The authors are grateful for the support of the Heising-Simons Foundation under Grant No.~2023--4612. This work is supported by the Fermilab Laboratory Directed Research and Development (LDRD) program. Fermilab operates under U.S. Department of Energy (DOE) contract No. DE-AC02-07CH11359. This work is partially supported by NASA APRA award No.\ 80NSSC22K1411. 

\addcontentsline{toc}{section}{References}
\bibliographystyle{JHEP}
\bibliography{references/books, references/skipperCCD, references/darkmatter}

\end{document}